\documentclass[CEJP,DVI,a4paper,final]{cej}
\usepackage{t1enc}
\usepackage{amssymb}
\usepackage{amsmath}
\usepackage{graphicx}
\usepackage{textcomp}
\usepackage{bm}

\renewcommand{\vec}[1]{\boldsymbol{#1}}

\newcommand{\bkt}[1]{\left({#1}\right)}
\newcommand{\abs}[1]{\left|{#1}\right|}
\newcommand{\od}[1]{\!\bkt{#1}}
\newcommand{\rt}{\od{\vec r,t}}
\newcommand{\e}{\vec e}
\newcommand{\f}{\vec f}
\newcommand{\fH}{\vec f^{_H}}

\newcommand{\ket}[1]{\left|{#1}\right>}
\newcommand{\bra}[1]{\left<{#1}\right|}
\newcommand{\imag}[1]{\mathrm{Im}\left\{{#1}\right\}}
\newcommand{\real}[1]{\mathrm{Re}\left\{{#1}\right\}}

\newcommand{\B}{\mathcal{B}}
\newcommand{\E}{\mathcal{E}}

\newcommand{\dip}{\mathrm{d}}
\newcommand{\tF}{\tilde F}

\newcommand{\aH}{a^\dagger}

\newcommand{\I}{\mathrm{i}}

\newcommand{\id}[3][)]{
\if #1)
\int\!\mathrm{d}^{#2}{#3}\,%
\else
\int_{#1}\!\mathrm{d}^{#2}{#3}\,%
\fi
}
\newcommand{\lid}[3]{\int_{#2}^{#3}\!\mathrm{d}{#1}\,}

\newcommand{\diff}[3][)]{
\if #1)
\frac{d{#2}}{d{#3}}%
\else
\frac{d^{#1}{#2}}{d^{#1}{#3}}%
\fi
}
\newcommand{\pdiff}[3][)]{
\if #1)
\frac{\partial {#2}}{\partial {#3}}%
\else
\frac{\partial^{#1}{#2}}{\partial^{#1}{#3}}%
\fi
}

\newcommand{\pol}{\epsilon}
\newcommand{\spol}[1][)]{
\if #1)
\gamma_{\pol}%
\else
\gamma_{#1}%
\fi
}
\newcommand{\TE}{\mathrm{TE}}
\newcommand{\TM}{\mathrm{TM}}
\newcommand{\RM}{{\mathrm{RM}}}
\newcommand{\SM}{{\mathrm{SM}}}
\newcommand{\GM}{{\mathrm{GM}}}
\newcommand{\RSM}{{\mathrm{RSM}}}

\newcommand{\pval}{\mathcal{P}}

\newcommand{\hatHam}{\hat H}
\newcommand{\hatHem}{\hatHam_{\mathrm{em}}}
\newcommand{\hatHint}{\hatHam_{\mathrm{int}}}
\newcommand{\hatHm}{\hatHam_{\mathrm{at}}}

\newcommand{\Dirac}[2][)]{
\if #1)
\delta\od{#2}%
\else
\delta^{({#1})}\od{#2}%
\fi
}
\newcommand{\trdelta}[2]{\delta_{\varepsilon\perp}^{#1}\od{#2}}

\newcommand{\locf}{\mathcal{L}_\mathrm{diel}}
\newcommand{\str}[1]{S{#1}}

\newcommand{\pbkt}[3][)]{
\if #1)
\left\{{#2},{#3}\right\}%
\else
\left\{{#2},{#3}\right\}_{\mathrm{#1}}%
\fi
}
\newcommand{\kom}[3][)]{
\if #1)
\left[{#2},{#3}\right]%
\else
\left[{#2},{#3}\right]_{\mathrm{#1}}%
\fi
}

\newlength{\sumwidth}%
\newcommand{\sumint}[1][]{%
{\centering%
\sum%
\settowidth{\sumwidth}{$\sum$}
\hspace{-\sumwidth}%
\settowidth{\sumwidth}{$\int$}
\hspace{-.5\sumwidth}%
\int}_{#1}%
}

\title{Calculation of atomic spontaneous emission rate in 1D finite photonic crystal with defects}

\articletype{Research Article}

\author{Adam Rudzi{\'n}ski\inst{1}\email{arudzins@poczta.onet.pl},
        Pawe\l{} Szczepa{\'n}ski\inst{1,2}}

\institute{
     \inst{1} Institute of Microelectronics and Optoelectronics, Warsaw University of Technology,\\
     ul.~Koszykowa~75, 00-665 Warszawa, Poland\\
     \inst{2} National Institute of Telecommunications,
     ul.~Szachowa~1, 00-894 Warszawa, Poland
          }

\abstract{
We derive the expression for spontaneous emission rate in finite one-dimensional photonic crystal
with arbitrary defects using the effective resonator model to describe electromagnetic field distributions
in the structure. We obtain explicit formulas for contributions of different types of modes,
i.e.~radiation, substrate and guided modes. Formal calculations are illustrated with a few numerical
examples, which demonstrate that the application of effective resonator model simplifies interpretation
of results.
}


\begin{document}
\maketitle

\section{Introduction}
It is a long known fact, that interaction of a system with electromagnetic radiation is affected not
only by the system itself, but also by the environment in which it is situated. Placement of an atom
or a molecule in a~cavity leads to many effects not present in unbounded vacuum, which are the subject
of studies within the area of so-called cavity QED -- for a review see e.g. \cite{Bialynicka96}.
In particular, one of these phenomena is the modification of spontaneous emission rate, predicted
by Purcell \cite{Purcell46} and later verified in many experiments, e.g. \cite{Goy83, deMartini91, Nishioka93, Tocci96}. This fact has
considerable consequences for contemporary science and technology, therefore it is an important
object of research. Because spontaneous emission is a quantum effect, for its modeling it is
necessary to employ a formalism based on quantization of electromagnetic field. A theoretical description
suitable for the treatment of spontaneous emission in dielectric structures, followed in many papers,
has been discussed by Glauber and Lewenstein \cite{Glauber91}.

Another fact, that has been recognized over a hundred years ago and described by Lord Rayleigh
\cite{Strutt87}, is that periodically arranged medium has peculiar properties, with the most interesting
one being the existence of band gaps -- frequency ranges in which no propagating waves exist. In photonics,
periodic dielectric materials with band gaps for electromagnetic waves (photonic band gaps) are called
photonic crystals. The simplest of this kind of structures are one-dimensional photonic crystals, built
of alternating layers with two different refractive indices and widths (and, in fact, can be considered
a particular type of planar multilayer waveguide). They are a particularly good subject of study, because
one-dimensional multilayers can be described analytically and are significant from the practical point of view.
As it has been shown, by proper choice of parameters it is possible to obtain a band gap for all directions of propagation \cite{Winn98, Chigrin99, Lee03},
but even without a~full band gap, one-dimensional periodical lattice allows for a significant modification
of emission of light and therefore is useful for construction of devices such as light emitting diodes
or lasers \cite{Tocci96, Bastonero99, Bugajski02}.
To understand their operation, and to be able to
design them with care for various details, it is important to properly describe the impact of the structure
on the process of emission of light.

A perfect photonic crystal, which is strictly periodic structure, consists of an infinite number
of elementary cells (i.e. groups of layers, which appear in the same sequence in every period)
and can be modeled with the use of Floquet-Bloch theorem, fixing the form of modes of the structure.
A recent example of this treatment is \cite{Sanchez05}, in which spontaneous emission has been described with
a classically derived formula. However, a~practical realization of a photonic crystal is inevitably
of finite size, thus, in a more realistic models, the structure must be treated rather as a~planar multilayer
waveguide. This approach makes it possible to account for defects (of width or refractive index)
of the structure as well. Just as in a perfect photonic crystal, it is possible to use a~classical
description of spontaneous emission in a~planar waveguide, e.g. \cite{Lukosz80}. One has to note, though,
that what is actually calculated this way, is radiation of a classical electric dipole, and only comparison
with a result from quantum theory allows to establish its relation to spontaneous emission. Hence, such
treatment is rather just a limited reconstruction of a specific formula, and not a relevant, general theory
in any way. A proper, versatile description of spontaneous emission has to be established in the quantum
framework, like \cite{Glauber91}.
For a multilayer waveguide, a particular case of which is a finite one-dimensional photonic crystal,
one cannot use the Floquet-Bloch theorem and has to choose modes of the structure in a~different way.
A basis suitable for a single interface between two media has been proposed by Carniglia and Mandel \cite{Carniglia71}.
In their proposition, the set of modes is constructed starting with plane waves propagating
towards the structure (sweeping through all frequencies and angles of incidence), thus it can be referred
to as a set of so-called incoming modes. Each incident plane wave has to be accompanied by respective
reflected and transmitted waves, and such triplet constitutes a field distribution in the structure.
Modes of this form are particularly simple, form orthogonal \cite{Hammer03}
and complete \cite{Bialynicki72} set. There also exists a modification of this model, in which
instead of waves propagating towards the interface, waves propagating away from the interface are used
\cite{Zakowicz02}, so-called outgoing modes.
With this choice there is always only one emitted plane wave carrying power out in a chosen direction,
and this set is considered favorable for this kind of calculations,
e.g. in \cite{Creatore08}. For a multilayer structure, it is necessary to supplement these models with guided modes,
calculated for the layer with the highest refractive index \cite{Rigneault96},
and radiation modes can be redefined as combinations of the incoming or outgoing modes, for which no power flow
in transverse direction occurs, what assures their orthogonality \cite{Wang97}.
Field distributions of modes can be conveniently found with the help of translation matrix method \cite{Yeh77, Yeh88}.

For all these kinds of modes, it is a common feature, that a kind of source of radiation exciting a mode is situated outside
the structure, what seems not to be particularly suitable for description of emission from a~system
located inside the structure, because it leads to treatment of the structure as a whole block, instead of distinguishing
particular locations inside, which may have significantly different properties.
However, it is possible to find field distributions starting from
a plane wave emerging in one of the layers. This approach, described or employed e.g. in \cite{Benisty98, Muszalski02},
requires to sum up all the reflected waves appearing in the layer, which resembles a~text-book exercise
of calculating transmission through a FP resonator. The difference is that in the case of source inside
the layer, reflections occur at both interfaces, thus there are twice as much waves to sum.
Modes obtained with this method are indexed by wave vectors and polarizations,
and naturally split into radiation, substrate and guided modes.
This construction of modes is also the starting point of the effective resonator model
\cite{ProcSPIE, APPA111, APPA112495, APPA112505, APPA115660}, which allows
a very clear physical interpretation of obtained results. The model is not just limited to calculating
distributions of electromagnetic field. Its main advantage is the definition of a quantity called mode spectrum,
which can be used to quickly and conveniently analyze modes and properties of the structure.
Assuming, that the interfaces between layers are perpendicular to $z$ axis, mode spectrum of a specific layer is defined as
\begin{equation}
\rho_\pol\od{\vec k} = \frac{1}{8\pi^3}\frac{1-\abs{r_Lr_R}^2}{\abs{1 - r_Lr_R\exp\od{2\I k_zL_z}}^2},
\end{equation}
where $\vec k$ is wave vector and $\pol$ -- polarization of the mode, $r_L$ and $r_R$ are reflection
coefficients of the stacks of layers on both sides of the layer with the source (i.e. subscript $L$ stands for ``left''
and $R$ -- for ``right'', these coefficients are functions of polarization and wave vector, but for
clarity of the expression it is not explicitly denoted), $k_z$ is the $z$ component of the wave vector
and $L_z$ is the layer's width. Similar expressions have already appeared in the literature, but they have never
been given such an important physical meaning. Mode spectrum is a quantity directly related to density of states and spontaneous emission rate,
resembling quality factor of a~cavity, which also takes
into account interference between plane waves composing a~mode. If the quality factor for a mode is high,
then the value of mode spectrum is high when the interference is constructive, or low, if the interference
is destructive. In particular, if mode spectrum is equal to zero, this means, that the field of the mode
is completely extinguished in the cavity. For modes, for which the quality factor is low, the value
of mode spectrum is close to free space value
\begin{equation}
\rho^\mathrm{fs}=\frac{1}{8\pi^3}.
\end{equation}

In this paper, we derive the expression for the spontaneous emission rate in a one-dimensional photonic
crystal with arbitrary defects of width or refractive index of any layers, split into explicit
contributions from different kinds of modes, i.e. radiation, substrate and guided modes,
obtained with the help of the effective resonator model.
It is shown, that mode spectrum characterizes spontaneous emission rates inherent for each layer. It
allows for an easy investigation of properties of the structure, in particular containing defects,
either introduced intentionally \cite{OQEL} or random fabrication impurities, which have negative influence
on the structure \cite{Eurocon}. Calculations conducted with effective resonator model lead to description
of spontaneous emission which is particularly convenient for research on impact of the structure on
the process or dealing with practical tasks, where a~proper design of the structure is necessary.
Obtained results can be used for a~detailed study on how a multilayer environment affects spontaneous emission
decay, first of all to identify the modes which contribute the most in particular layers, what is not straightforward
with other construction of modes. Effective resonator model is completely analytic, thus, in principle, it allows
to obtain analytic expressions describing various aspects of the phenomenon
or could be used as a basis for semi-analytic or approximate calculations in more complicated
structures. The paper is organized as follows. In Sec.~\ref{sec_strnot}, we define the considered structure,
introduce dimensionless parameters, particularly suitable for analysis of its properties, and
briefly review a few properties of modes in the effective resonator model. Next, in Sec.~\ref{sec_hamiltonian},
we describe the quantum formalism adopted for calculations, which we employ in Sec.~\ref{sec_ser} to obtain
the expression for spontaneous emission rate. Contributions from different kinds of modes are the subject
of calculations in Sec.~\ref{sec_calc}. We provide a few exemplary results in Sec.~\ref{sec_results}
and Sec.~\ref{sec_summary} summarizes the paper.

\section{Considered structure and adopted notation}\label{sec_strnot}
\begin{figure}[tb]
\begin{center}
\includegraphics[width=.45\textwidth]{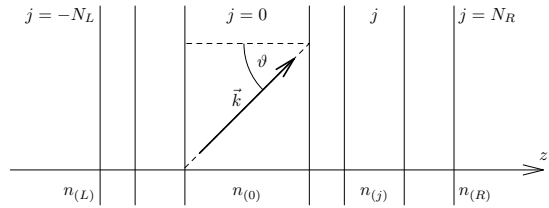}
\caption{Sketched structure of dielectric multilayer waveguide.}\label{fig_dmw}
\end{center}
\end{figure}
We consider a structure of finite one-dimensional photonic crystal, which is a particular case of
a dielectric multilayer waveguide, sketched in Fig.~\ref{fig_dmw}. We define the $z$ axis perpendicular
to interfaces between dielectric layers, which we assume to be made of lossless, isotropic dielectrics.
The refractive index of the structure is then given by
\begin{equation}
n\od{\vec r}=\sum_j\chi_j\od{z} n_{(j)},
\end{equation}
where $j$ indexes layers, $n_{(j)}$ is the refractive index of the $j$th layer, and $\chi_j\od{z}$
the characteristic function of the $j$th layer: $\chi_j\od{z}=1$ for $z$ in the $j$th layer and
$\chi_j\od{z}=0$ for $z$ from outside of it. Index $j$ runs from $-N_L\leq 0$ to $N_R\geq 0$ and we put
$z=0$ at the left boundary of the layer for which we choose $j=0$. We treat the media outside as
layers with $j=-N_L$ and $j=N_R$, but additionally assume, that $n_{(R)}\leq n_{(L)}$  (in subscripts
and superscripts we use abbreviations $R\equiv N_R$ and $L\equiv -N_L$) -- this is just the matter of $z$
axis orientation, therefore it does not affect generality of the model or presented calculations.
We denote the width of the $j$th layer by $L_z^{(j)}$, using for the $j=0$ layer $L_z\equiv L_z^{(0)}$,
and assume that layers extend to infinity in $x$ and $y$ directions. For one-dimensional photonic crystal
of our concern
\begin{subequations}
\begin{equation}
n_{(j)}=
\begin{cases}
n_1 & \text{for even }j,\\
n_2 & \text{for odd }j,
\end{cases}\quad
\end{equation}
and
\begin{equation}
L_z^{(j)}=
\begin{cases}
L_1 & \text{for even }j,\\
L_2 & \text{for odd }j,
\end{cases}
\end{equation}
\end{subequations}
except for defected layers and the surrounding media, for which the general notation $n_{(j)}$
and $L_z^{(j)}$ has to be kept.

Modes of the structure are indexed with wave vector $\vec k$ and polarization $\pol=\TE,\TM$. To each
of them there are bound electric and magnetic field distributions, $\f_{k\pol}\od{\vec r}$
and $\fH_{k\pol}\od{\vec r}$. Different layers are equipped with different sets of modes, in the paper
we always refer to modes of the layer for which $j=0$, thus we obtain results relevant for this
particular layer (but chosen arbitrarily). Construction of these modes is discussed with details
in \cite{APPA111, APPA112495, APPA112505, APPA115660}, here we wish to briefly review only a few
of their properties.

Electric and magnetic fields radiated by a source inside the $j=0$ layer can be written in terms of
field distributions $\f_{k\pol}$ and $\fH_{k\pol}$:
\begin{subequations}\label{EHcl}
\begin{align}
\vec E\rt\equiv & \sumint[\vec k,\pol] \E_{k\pol}\od{t}\f_{k\pol}\od{\vec r}\label{sumint_def}\\
         =&\sum_\pol\id[\RSM]{3}{k}\E_{k\pol}\od{t}\f_{k\pol}\od{\vec r}\nonumber\\
          &+\sum_\pol\id{2}{k_\parallel}\sum_{a\in\GM\od{k_\parallel,\pol}}\E_{k_a\pol}\od{t}\f_{k_a\pol}\od{\vec r},\nonumber\\
\vec H\rt=&\sumint[\vec k,\pol] \B_{k\pol}\od{t}\fH_{k\pol}\od{\vec r},
\end{align}
\end{subequations}
where $\E_{k\pol}\od{t}$ and $\B_{k\pol}\od{t}$ are time-dependent amplitudes, $\RSM$ denotes subset
of radiation and substrate modes, and $\GM\od{k_\parallel,\pol}$ -- discrete subset of guided modes
with given $k_\parallel\equiv\sqrt{k_x^2+k_y^2}$ and polarization $\pol$. For guided modes
the integration is performed over components $k_x$ and $k_y$, while the whole wave vector of the
$a$th guided mode is $\vec k_a\equiv k_x\e_x+k_y\e_y+k_{z,a}\e_z$. Decomposition of modes into radiation,
substrate and guided modes is based on a~standard criterion of total reflection, we precisely define
each of these subsets in Sec.~\ref{sec_calc}. To improve clarity, we use the ``sum-integral'' symbol,
which combines integration over components of wave vector and summation over discrete guided modes
-- expression \eqref{sumint_def} can be considered its definition.

Every mode can be identified with its electric field distribution, because magnetic field distributions
can be obtained from the relation:
\begin{equation}
\fH_{k\pol}=\frac{\nabla\times\f_{k\pol}}{\I\mu_0\omega_k},
\end{equation}
where
\begin{equation}
\omega_k = \frac{c}{n_{(0)}}\sqrt{\vec k^2}
\end{equation}
is the angular frequency of the mode. The last expression is the well known dispersion relation for a plane wave,
which is a condition the plane wave must satisfy to be a valid mathematical solution of Maxwell equations,
in which $\omega_k$ and components of $\vec k$ can be in general complex numbers. Further restrictions can be
settled by requirement, that the field must be physical, i.e. must not ``explode'' in infinity, meaning that the exponents
$\exp\od{\I\vec k\cdot\vec r}\exp\od{-\I\omega_k t}$ should either oscillate or fade.
An obvious notion is that $\omega_k\in\mathbb{R}$, because of the time uniformity, similarly
$k_x,k_y\in\mathbb{R}$, because the considered structure is uniform in $x$ and $y$ directions,
but spatial dependence in $z$ direction is more complicated. The field is always physical if $k_z\in\mathbb{R}$,
but in general it can be as well for some imaginary values $k_z\in\I\mathbb{R}$,
if the refractive index of the layer is sufficiently small (the field has to be oscillatory in another layer, to connect
two evanescent waves fading in opposite directions). However, in this paper we wish to concentrate only on modes
with real wave vectors $\vec k\in\mathbb{R}^3$, i.e. our considerations are limited to layers with sufficiently high
refractive index. The reason is that for a single non-uniform plane wave with imaginary $k_z$
there is no flow of energy in $z$ direction, thus, a natural choice of waves on an interface between two layers
does not contain a~reflected wave. As our definition of mode spectrum is based on reflection coefficients,
it cannot be naturally applied for non-uniform plane waves, and we wish to show in the paper that
this quantity is important for spontaneous emission. Therefore, from now on we assume $k_z\in\mathbb{R}$.

Modes $\f_{k\pol}$ are orthonormal, and their orthonormalization
rules are:
\begin{subequations}
\begin{equation}
\id{3}{r}\varepsilon\od{\vec r}\f_{q\lambda}^*\od{\vec r}\cdot\f_{k\pol}\od{\vec r}=\delta_{\pol\lambda}\Dirac{\vec k-\vec q},
\end{equation}
for radiation and substrate modes, while for guided modes:
\begin{multline}
\id{3}{r}\varepsilon\od{\vec r}\f_{q_b\lambda}^*\od{\vec r}\cdot\f_{k_a\pol}\od{\vec r}=\\\delta_{\pol\lambda}\Dirac{k_x-q_x}\Dirac{k_y-q_y}\delta_{ab}.
\end{multline}
\end{subequations}
In the considered case, relative electric permittivity $\varepsilon\od{\vec r}=n^2\od{\vec r}$.
Modes $\f_{k\pol}$ can be used to represent the generalized transverse delta function $\trdelta{ij}{\vec r,\vec r'}$,
which is a tensor conserving transverse fields. It is defined by the relation \cite{Glauber91}
\begin{equation}
\id{3}{r'}\varepsilon\od{\vec r'}\trdelta{ij}{\vec r,\vec r'}f_{k\pol}^j\od{\vec r'}=f_{k\pol}^i\od{\vec r}
\end{equation}
(in this, and the remaining expressions in the paper, we use the summation convention for upper indices).
The generalized transverse delta can then be represented as:
\begin{equation}
\trdelta{ij}{\vec r,\vec r'} = \sumint[\vec k,\pol] f_{k\pol}^i\od{\vec r}f_{k\pol}^{j*}\od{\vec r'}.
\end{equation}
This function has two important properties, it is real:
\begin{equation}
\trdelta{ij*}{\vec r,\vec r'}=\trdelta{ij}{\vec r,\vec r'}
\end{equation}
and
\begin{equation}
\trdelta{ij}{\vec r,\vec r'}=\trdelta{ji}{\vec r',\vec r}.
\end{equation}
These identities can be easily proven using the relation $\f_{k\pol}^*=-\spol\f_{-k\pol}$, with
\begin{equation}
\spol=
\begin{cases}
1,&\text{for }\pol=\TE,\\
-1,&\text{for }\pol=\TM,
\end{cases}
\end{equation}
which is satisfied by all modes, and $s_{k\pol}\f_{k_*\pol}=\f_{k\pol}$, satisfied by substrate
and guided modes, for which coefficient $s_{k\pol}=r_R\exp\od{2\I k_zL_z}$ \cite{APPA115660}.

Elementary cell of the considered photonic crystal consists of two layers, the first one with width
$L_1$ and refractive index $n_1$, and the second, with width $L_2$ and refractive index $n_2$.
Optical widths of these layers are defined as $\Lambda_1=n_1 L_1$, $\Lambda_2=n_2 L_2$, in general,
for the $j$th layer $\Lambda^{(j)}=n_{(j)} L_z^{(j)}$. Optical width of the whole elementary cell is then
$\Lambda=\Lambda_1+\Lambda_2$. It is known, that properties of photonic crystals depend on proportions
of the structure, thus, to investigate its properties
it is convenient to use parameters which are dimensionless, and we use $\Lambda$ as a unit of length
in the construction of such set. To describe the structure, we use refractive indices, normalized
optical widths $\Lambda_i/\Lambda$ (in general $\Lambda^{(j)}/\Lambda$), normalized position in
the $j=0$ layer $\zeta=n_{(0)} z/\Lambda^{(0)}$ and normalized frequency $\varTheta=f/f_0$, where
$f_0=c/\Lambda$. To obtain more compact expressions, instead of the angle of incidence $\vartheta$
we use its cosine $\eta=\cos\!\vartheta$. All expressions of our interest can be written in terms of
these parameters, e.g.:
\begin{subequations}
\begin{align}
k_z z&=2\pi\eta\varTheta\zeta\frac{\Lambda^{(0)}}{\Lambda},\\
k_z^{(j)}L_z^{(j)}&=\pm 2\pi\varTheta\sqrt{1-\frac{1-\eta^2}{n_{(j)}^2/n_{(0)}^2}}\frac{\Lambda^{(j)}}{\Lambda},\\
\abs{\frac{k_z^{(j)}}{k_z}}&=\sqrt{\abs{1+\frac{n_{(j)}^2/n_{(0)}^2-1}{\eta^2}}}.
\end{align}
\end{subequations}

\section{Hamiltonian of the system}\label{sec_hamiltonian}
Electromagnetic field and a quantum system (e.g. an atom) interacting with each other can be described
with the hamiltonian consisting of three parts:
\begin{equation}
\hatHam=\hatHem+\hatHint+\hatHm.
\end{equation}
$\hatHem$ is the hamiltonian of electromagnetic field in inhomogeneous dielectric \cite{Glauber91}:
\begin{multline}
\hatHem=\\\frac{1}{2}\id{3}{r}:\!\bkt{\varepsilon_0\varepsilon\od{\vec r}\vec E^2\od{\vec r}+\frac{\bkt{\nabla\times\vec A\od{\vec r}}^2}{\mu_0}}\!:,
\end{multline}
where colons denote normal ordering, $\vec E$ and $\vec A$ are electric field and vector potential
operators. $\hatHm$ is the hamiltonian of the quantum system. Because we are interested in the influence
of the structure, we use the simplest possible expression, which for an electronic system, an atom or
a molecule, in the infinite nuclear mass approximation (so-called Born-Oppenheimer approximation
for molecules) is the non-relativistic hamiltonian
\begin{equation}
\hatHm=\sum_a \frac{\vec p_a^{\,2}}{2m}+\hat V,
\end{equation}
where $m$ is the electron's mass, $\vec p_a$ is the $a$th electron's momentum operator and $\hat V$ is the potential operator,
describing interactions of electrons and atomic nuclei. Formally, if $\ket{\Psi_i}$ denote
eigenstates of $\hatHm$ with energies $E_i$, it can be written
\begin{equation}
\hatHm=\sum_i E_i\ket{\Psi_i}\bra{\Psi_i}.
\end{equation}
$\hatHint$ is the interaction hamiltonian, which can be derived from the hamiltonian $\hatHm$
through the minimal coupling $\vec p_a\rightarrow \vec\pi_a=\vec p_a-e\vec A_a$, with $e<0$ being
the electron's charge and $\vec A_a\equiv\vec A\od{\vec r_a}$ denoting vector potential in the
position of the $a$th electron. In vacuum
\begin{equation}
\hatHm+\hatHint^\mathrm{vac}=\sum_a \frac{\vec\pi_a^2}{2m}+\hat V.
\end{equation}
The $\vec A_a^{\,2}$ term emerging from $\vec\pi_a^2$ does not contribute to spontaneous emission, at least directly,
because it contains no atomic operators and we neglect this term (what is in fact a usual procedure).
Because $\kom{r_a^i}{p_b^j}=\I\hbar\delta_{ab}\delta^{ij}$, momentum $\vec p_a$ can be expressed as
\begin{equation}
p_a^i=\frac{m}{\I\hbar}\kom{r_a^i}{\hatHm}.
\end{equation}
We wish to concentrate on interaction of electrons localized in a small volume near an atomic nucleus
(or nuclei in case of a small molecule), what is the case for e.g. dopant atoms in solid state lasers,
where emission is related to transition between two atomic levels. It is
possible to study transitions of non-localized electrons, like in conduction band of bulk semiconductor
(see e.g. \cite{Hooijer01}), but that is not the case we are going to take into consideration in this
paper. In the case of our interest, because optical wavelengths, which are of concern in case of light
emitting devices, are much longer than the radius of an atom, the interaction Hamiltonian can be simplified
by application of the dipole approximation: $\vec A_a\approx\vec A\od{\vec r_0}$, where $\vec r_0$
is the position of the system (understood as e.g. position of the center of mass). Then
\begin{equation}
\hatHint^\mathrm{vac}=\frac{\I}{\hbar}A^i\od{\vec r_0}\kom{\dip^i}{\hatHm},
\end{equation}
where electric dipole moment $\vec\dip=\sum_a e\vec r_a$. In dielectric, the dipole moment interacts not
with the macroscopic field, which is found as solution of Maxwell equation, but with microscopic
local field \cite{Glauber91}. This effect can be accounted for by inclusion of field
enhancement factor $\locf$, usually defined as the ratio of local and macroscopic electric field,
but since the relation between electric field and vector potential is linear, it can be used with vector
potential as well. Thus, the interaction hamiltonian in dielectric medium becomes
\begin{equation}
\hatHint=\frac{\I}{\hbar}\locf A^i\od{\vec r_0}\kom{\dip^i}{\hatHm}.\label{Hintdiel}
\end{equation}
At this moment it is not necessary to define $\locf$ explicitly by an expression, and because its value
depends on applied theory, we proceed further using the general symbol. The interaction hamiltonian
obtained in the presented derivation is similar to the $\vec E\cdot\vec\dip$ form, but not
equivalent. This form could be reconstructed if $\kom{\vec\dip}{\hatHint}=0$, because the commutator
in \eqref{Hintdiel} would denote the time derivative, which could be transferred onto the vector potential,
turning it into electric field. However, it is easy to check, that with
the dipole approximation $\kom{\vec\dip}{\hatHint}\approx-\I\hbar\frac{Ne^2}{m}\locf\vec A\od{\vec r_0}$,
where $N$ stands for the number of electrons. Hence, we use the expression \eqref{Hintdiel}
for the interaction hamiltonian.

In dielectric, vector potential and electric field operators obey the commutation rule \cite{Glauber91}:
\begin{equation}
\kom{A^i\od{\vec r}}{E^j\od{\vec r'}}=-\frac{\I\hbar}{\varepsilon_0}\trdelta{ij}{\vec r,\vec r'}
\end{equation}
and can be in the usual way expanded into the effective resonator modes:
\begin{widetext}
\begin{subequations}
\begin{align}
\vec A\od{\vec r}=&\sumint[\vec k,\pol] \sqrt{\frac{\hbar}{2\varepsilon_0\omega_k}}\bkt{a_{k\pol}\f_{k\pol}\od{\vec r}+\aH_{k\pol}\f_{k\pol}^*\od{\vec r}},\\
\vec E\od{\vec r}=&\I\sumint[\vec k,\pol] \sqrt{\frac{\hbar\omega_k}{2\varepsilon_0}}\bkt{a_{k\pol}\f_{k\pol}\od{\vec r}-\aH_{k\pol}\f_{k\pol}^*\od{\vec r}},
\end{align}
\end{subequations}
where $a_{k\pol}$ and $\aH_{k\pol}$ are the photon annihilation and creation operators, obeying the bosonic
commutation rules:
\begin{subequations}
\begin{align}
&\kom{a_{k\pol}}{a_{q\lambda}}=\kom{\aH_{k\pol}}{\aH_{q\lambda}}=0,\\
&\kom{a_{k\pol}}{\aH_{q\lambda}}=\delta_{\pol\lambda}\Dirac{\vec k-\vec q} \quad\text{(radiation and substrate modes)},\\
&\kom{a_{k_a\pol}}{\aH_{q_b\lambda}}=\delta_{\pol\lambda}\Dirac{k_x-q_x}\Dirac{k_y-q_y}\delta_{ab}\quad\text{(guided modes)}.
\end{align}
\end{subequations}
Hamiltonian of electromagnetic field obtains then the form characteristic to an ensemble of harmonic oscillators:
\begin{equation}
\hatHem=\sumint[\vec k,\pol] \hbar\omega_k\hat N_{k\pol},
\end{equation}
where $\hat N_{k\pol}=\aH_{k\pol}a_{k\pol}$
is the photon number operator with eigenvectors $\ket{\ldots n_{k\pol}\ldots}$,
where $n_{k\pol}=0,1,2,\ldots$ and:
\begin{subequations}
\begin{align}
a_{k\pol}\ket{\ldots n_{k\pol}\ldots}&=\sqrt{n_{k\pol}}\ket{\ldots \bkt{n_{k\pol}-1}\ldots},\\
\aH_{k\pol}\ket{\ldots n_{k\pol}\ldots}&=\sqrt{n_{k\pol}+1}\ket{\ldots \bkt{n_{k\pol}+1}\ldots}.
\end{align}
\end{subequations}
\end{widetext}

\section{Spontaneous emission}\label{sec_ser}
Spontaneous emission is a process in which a relaxing atom emits a single photon.
Thus, for the description of this phenomenon, the Hilbert space of $\hatHm$
can be restricted to only two eigenstates $\ket{\Psi_0}$ and $\ket{\Psi_1}$
(we assume that $E_1>E_0$), and the Hilbert space of $\hatHem$ to eigenstates $\ket{0}$
-- with no photons, and $\ket{1_{k\pol}}$ -- with one photon in the mode with wave vector
$\vec k$ and polarization $\pol$. For the whole system, the excited state $\ket{\Psi_1}$
is accompanied by the state with no photons $\ket{0}$ and the lower state $\ket{\Psi_0}$
-- by field with one photon. The proper state vector is then:
\begin{equation}
\ket{\psi\od{t}} = C_{0}\od{t}\ket{\Psi_1}\ket{0} + \sumint[\vec k,\pol] C_{k\pol}\od{t}\ket{\Psi_0}\ket{1_{k\pol}}.
\end{equation}
The initial conditions adequate for spontaneous emission are:
\begin{eqnarray}
C_0\od{t}=1,&&C_{k\pol}\od{t}=0.
\end{eqnarray}
In the two-level case, states $\ket{\Psi_0}$ and $\ket{\Psi_1}$ can be chosen
so that the transition dipole moment $\vec\dip_{10}=\bra{\Psi_1}\vec\dip\ket{\Psi_0}\in\mathbb{R}^3$. In this paper
we calculate the spontaneous emission rate following the procedure described in \cite{Meystre91}.
After substitutions:
\begin{subequations}
\begin{align}
C_0\od{t}&=b_0\od{t}\exp\od{-\frac{\I E_1 t}{\hbar}},\\
C_{k\pol}\od{t}&=b_{k\pol}\od{t}\exp\od{-\I\frac{\hbar\omega_k+E_0}{\hbar}t},
\end{align}
\end{subequations}
where $b_0\od{t}$ and $b_{k\pol}\od{t}$ are the probability amplitudes in the interaction picture,
the equations of motion become:
\begin{subequations}
\begin{align}
\diff{b_0}{t}=&\sumint[\vec k,\pol] g_{k\pol}b_{k\pol}\od{t}\mathrm{e}^{\I\bkt{\Omega_{10}-\omega_k}t},\\
\diff{b_{k\pol}}{t}=&-g_{k\pol}^*b_0\od{t}\mathrm{e}^{\I\bkt{\omega_k-\Omega_{10}}t},\label{dotbkpol}
\end{align}
\end{subequations}
where angular frequency of the atomic transition $\Omega_{10}=\bkt{E_1-E_0}/\hbar$
and
\begin{equation}
g_{k\pol}=\frac{\bra{0}\bra{\Psi_1}\hatHint\ket{\Psi_0}\ket{k\pol}}{\I\hbar}.
\end{equation}
In this case, if $\vec r_0$ denotes the position of the quantum system:
\begin{equation}
g_{k\pol}=-\frac{\locf\Omega_{10}\vec\dip_{10}\cdot\f_{k\pol}\od{\vec r_0}}{\sqrt{2\varepsilon_0\hbar\omega_k}}.
\end{equation}
Symmetry of multilayer structure allows to put $\vec r_0=z_0\vec e_z$.

Formal solution of \eqref{dotbkpol} is:
\begin{equation}
b_{k\pol}\od{t}=-g_{k\pol}^*\lid{\tau}{0}{t}b_0\od{\tau}\mathrm{e}^{\I\bkt{\omega_k-\Omega_{10}}\tau},
\end{equation}
therefore:
\begin{widetext}
\begin{equation}
\diff{b_0}{t}=-\sumint[\vec k,\pol] \abs{g_{k\pol}}^2\lid{\tau}{0}{t}b_0\od{\tau}\mathrm{e}^{\I\bkt{\Omega_{10}-\omega_k}\bkt{t-\tau}}.
\end{equation}
Assuming that $b_0$ varies slowly during the time interval from $0$ to $t$ and pulling it in front of the
time integral (Markov process approximation) one obtains:
\begin{equation}
\diff{b_0}{t}=-\sumint[\vec k,\pol] \abs{g_{k\pol}}^2 b_0\od{t}\lid{\tau}{0}{t}\mathrm{e}^{\I\bkt{\Omega_{10}-\omega_k}\bkt{t-\tau}}.
\end{equation}
\end{widetext}
In the limit $t\rightarrow\infty$ the integral over $\tau$ becomes:
\begin{multline}
\lim_{t\rightarrow\infty}\lid{\tau}{0}{t}\mathrm{e}^{\I\bkt{\Omega_{10}-\omega_k}\bkt{t-\tau}}=\\
\pi\Dirac{\Omega_{10}-\omega_k}-\pval\frac{\I}{\Omega_{10}-\omega_k},
\end{multline}
where $\mathcal{P}$ denotes principal value. The term containing the delta function describes the decay
process and can be used to determine the spontaneous emission rate, while the principal value term
results in line shift $\delta\omega$, which is not of our concern here. In the long time limit
\begin{equation}
\diff{b_0}{t}=\bkt{-\frac{\Gamma}{2}+\I\delta\omega} b_0\od{t},
\end{equation}
therefore
\begin{equation}
b_0\od{t}=\exp\od{-\frac{\Gamma t}{2}}\mathrm{e}^{\I\delta\omega t},
\end{equation}
where
\begin{equation}
\Gamma=\Gamma_\RM+\Gamma_\SM+\Gamma_\GM
\end{equation}
is the spontaneous emission rate. Contributions from different kinds of modes to $\Gamma$ are:
\begin{subequations}
\begin{align}
\Gamma_\RM=&\,2\pi\sum_\pol\id[\RM]{3}{k} \abs{g_{k\pol}}^2 \Dirac{\Omega_{10}-\omega_k},\\
\Gamma_\SM=&\,2\pi\sum_\pol\id[\SM]{3}{k} \abs{g_{k\pol}}^2 \Dirac{\Omega_{10}-\omega_k},\\
\Gamma_\GM=&\,2\pi\sum_\pol\id{2}{k_\parallel}\sum_{a\in\GM\od{k_\parallel,\pol}}\nonumber\\
           &\times\abs{g_{k_a\pol}}^2 \Dirac{\Omega_{10}-\omega_{k_a}}.
\end{align}
\end{subequations}
In case of free space, expression for $\Gamma$ leads to the well-known Weisskopf-Wigner rate:
\begin{equation}
\Gamma^\mathrm{fs}=\frac{\mu_0\Omega_{10}^3\dip_{10}^2}{3\pi\hbar c}.
\end{equation}
With the knowledge of $b_0$, solution of the equation of motion for $b_{k\pol}$ is:
\begin{equation}
b_{k\pol}\od{t}=g_{k\pol}^*\frac{1-\exp\od{\I\bkt{\delta\omega+\omega_k-\Omega_{10}}t-\frac{\Gamma t}{2}}}{\I\bkt{\delta\omega+\omega_k-\Omega_{10}}-\frac{\Gamma}{2}},
\end{equation}
thus, in the limit $t\rightarrow\infty$:
\begin{equation}
\abs{b_{k\pol}\od{t}}^2=\frac{\abs{g_{k\pol}}^2}{\bkt{\delta\omega+\omega_k-\Omega_{10}}^2+\frac{\Gamma^2}{4}}.\label{eq_absbkpol2}
\end{equation}

\section{Numerical calculations}\label{sec_calc}
The integration of $\abs{g_{k\pol}}^2$ necessary to calculate numerical value of $\Gamma$ is best performed in spherical
coordinates $\bkt{k,\vartheta,\varphi}$, where $\vartheta$ is the angle of incidence and $\varphi$
is the azimuthal angle. Results of integration over $\varphi$ can be obtained easily. Every $\f_{k\TE}$
lies within the $xy$ plane and can be written in function of $\varphi$:
\begin{equation}
\vec f_{k\TE}=-P_k\vec e_x\sin\!\varphi+P_k\vec e_y\cos\!\varphi,
\end{equation}
where
\begin{equation}
P_k=\vec e_y\cdot\f_{k\TE}\Big|_{\varphi=0}.
\end{equation}
Because the structure has rotational symmetry, axes $x$ and $y$ can be chosen in the way that
\begin{equation}
\vec\dip_{10} = \dip_\parallel\vec e_x+\dip_z\vec e_z.
\end{equation}
Then:
\begin{equation}
\lid{\varphi}{0}{2\pi}\abs{\vec\dip_{10}\cdot\f_{k\TE}}^2 = \pi\dip_\parallel^2\abs{P_k}^2.
\end{equation}
Every $\f_{k\TM}$ lies in the plane of incidence and can be written in function of $\varphi$:
\begin{equation}
\f_{k\TM}=Q_{k\parallel}\vec e_x\cos\!\varphi+Q_{k\parallel}\vec e_y\sin\!\varphi+Q_{k\perp}\vec e_z,
\end{equation}
with
\begin{subequations}
\begin{equation}
Q_{k\parallel}=\vec e_x\cdot\f_{k\TM}\Big|_{\varphi=0}
\end{equation}
and
\begin{equation}
Q_{k\perp}=\vec e_z\cdot\f_{k\TM}.
\end{equation}
\end{subequations}
Then:
\begin{equation}
\lid{\varphi}{0}{2\pi}\abs{\vec\dip_{10}\cdot\f_{k\TM}}^2=
\pi\dip_x^2\abs{Q_{k\parallel}}^2+2\pi\dip_z^2\abs{Q_{k\perp}}^2.
\end{equation}
Integration over $\vartheta$ has to be performed numerically.

\subsection{Radiation modes}
Radiation modes spread in the range of $0\leq\vartheta\leq\vartheta_S$ and $\pi-\vartheta_S\leq\vartheta\leq\pi$, where
\begin{equation}
\vartheta_S=\arcsin\frac{\min\left\{n_{(0)},n_{(L)},n_{(R)}\right\}}{n_{(0)}}.
\end{equation}
Contribution of radiation modes $\Gamma_\RM$ to spontaneous emission rate is:
\begin{widetext}
\begin{equation}
\Gamma_\RM=\frac{2\pi n_{(0)}^3\Omega_{10}^2}{c^3}\sum_\pol\lid{\vartheta}{0}{\vartheta_S}\sin\!\vartheta
\lid{\varphi}{0}{2\pi}\left[\abs{g_{k\pol}}^2+\abs{g_{k_*\pol}}^2\right]_{k=n_{(0)}\Omega_{10}/c}.
\end{equation}
In the second integral over $\varphi$ it is possible to change the variable to $\varphi+\pi$ and obtain
$\abs{g_{-k\pol}}$ instead of $\abs{g_{k_*\pol}}$. Then, because $\abs{g_{-k\pol}}=\abs{g_{k\pol}}$, using
the known results of integration over $\varphi$, after a few operations one arrives at
\begin{equation}
\Gamma_\RM=\frac{6\pi^3 n_{(0)}^3\locf^2\Gamma^\mathrm{fs}}{\dip_{10}^2}\lid{\eta}{\eta_S}{1}
\left[\dip_x^2\bkt{\abs{P_k}^2+\abs{Q_{k\parallel}}^2}+2\dip_z^2\abs{Q_{k\perp}}^2\right]_{\varTheta=\varTheta_{10}},\label{GammaRM}
\end{equation}
with normalized transition frequency:
\begin{equation}
\varTheta_{ij}=\frac{\Omega_{ij}}{2\pi f_0},
\end{equation}
$\eta_S=\cos\!\vartheta_S$ and
\begin{subequations}\label{PQ_RM}
\begin{align}
&\abs{P_k}^2 = \frac{ \rho_\TE\od{\vec k} }{ n_{(0)}^2 } \bkt{1 + \real{ \xi_\TE\od{\vec k} \mathrm{e}^{2\I k_z z_0} }},\\
&\abs{Q_{k\parallel}}^2 = \eta^2 \frac{ \rho_\TM\od{\vec k} }{ n_{(0)}^2 } \bkt{ 1-\real{\xi_\TM\od{\vec k}\mathrm{e}^{2\I k_z z_0}} },\\
&\abs{Q_{k\perp}}^2 = \bkt{1-\eta^2} \frac{ \rho_\TM\od{\vec k} }{ n_{(0)}^2 } \bkt{1 + \real{\xi_\TM\od{\vec k}\mathrm{e}^{2\I k_z z_0}} }.
\end{align}
\end{subequations}
For radiation modes
\begin{equation}
\xi_\pol\od{\vec k}=\frac{r_R^*\bkt{1-\abs{r_L}^2}\mathrm{e}^{-2\I k_zL_z}+r_L\bkt{1-\abs{r_R}^2}}{1-\abs{r_Lr_R}^2}.
\end{equation}
Coefficients \eqref{PQ_RM} can be easily calculated using explicit definitions of field distributions from \cite{APPA112495} or \cite{APPA115660}
(we do not recall them here, because it would be necessary to refer to a significant part
of calculations from \cite{APPA112495} in order to explain the symbols used in them),
knowing, that in those papers a few formulas for radiation modes can be significantly simplified, i.e.:
\end{widetext}
\begin{subequations}
\begin{align}
&F_{k\pol} = \frac{ n_{(0)}^2 }{ \rho_\pol\od{\vec k} },\label{F_RM}\\
&\tF_{k\pol} = \frac{n_{(0)}^2 \xi_\pol^*\od{\vec k}}{\rho_\pol\od{\vec k}},\\
&s_{k\pol} = \frac{ 1 - \sqrt{1 - \abs{\xi_\pol\od{\vec k}}^2} }{ \xi_\pol\od{\vec k} }.
\end{align}
\end{subequations}
In free space, where all the modes are of radiation type, \eqref{GammaRM} correctly reproduces the Weisskopf-Wigner
spontaneous emission rate, and in a multilayer structure, contribution of each radiation mode
is proportional to its mode spectrum.

\subsection{Substrate modes}
Substrate modes are found in the ranges of $\vartheta_S<\vartheta<\vartheta_G$ and $\pi-\vartheta_G<\vartheta<\pi-\vartheta_S$,
where
\begin{equation}
\vartheta_G=\arcsin\frac{\min\left\{n_{(0)},\max\left\{n_{(L)},n_{(R)}\right\}\right\}}{n_{(0)}}.
\end{equation}
However, these ranges contain the same modes \cite{APPA115660},
therefore they must not both enter into
the integration. Thus, the contribution to spontaneous emission rate is given by:
\begin{align}
\Gamma_\SM=&\,\frac{3\pi^3 n_{(0)}^3\locf^2\Gamma^\mathrm{fs}}{\dip_{10}^2} \lid{\eta}{\eta_G}{\eta_S}
\\&\times\left[\dip_x^2\bkt{\abs{P_k}^2+\abs{Q_{k\parallel}}^2}+2\dip_z^2\abs{Q_{k\perp}}^2\right]_{\varTheta=\varTheta_{10}},\nonumber
\end{align}
where $\eta_G=\cos\!\vartheta_G$. For substrate modes:
\begin{subequations}
\begin{align}
&\abs{P_k}^2 = \,2\frac{1+\real{\xi_{\TE}\od{\vec k}\exp\od{2\I k_z z_0}}}{F_{k\TE}},\\
&\abs{Q_{k\parallel}}^2 = \,2\eta^2\frac{1-\real{\xi_{\TM}\od{\vec k}\exp\od{2\I k_z z_0}}}{F_{k\TM}},\\
&\abs{Q_{k\perp}}^2 = \,2\bkt{1-\eta^2}\nonumber\\
                   &\quad\quad\quad\quad\quad\times\frac{1+\real{\xi_{\TM}\od{\vec k}\exp\od{2\I k_z z_0}}}{F_{k\TM}},
\end{align}
\end{subequations}
where $\xi_\pol\od{\vec k}=r_R^*\exp\od{-2\I k_zL_z}$, with the assumption that $n_{(R)}\leq n_{(L)}$,
and the coefficient $F_{k\pol}$ is given by \eqref{F_RM}, just like in case of radiation modes, thus, these expressions
are also proportional to mode spectrum.

\subsection{Guided modes}
Guided modes occur at discrete angles $\vartheta_a$ from the range
$\vartheta_G\leq\vartheta_a\leq\pi/2$ or $\pi/2\leq\vartheta_a\leq\pi-\vartheta_G$. Just like in the case
of substrate modes, guided modes with $\vartheta_a$ and $\pi-\vartheta_a$ (and the same polarization) are in
fact the same mode, therefore summation has to restrict to only one of these angles. Angles of incidence
of guided modes are determined by the relation
\begin{equation}
\mathrm{e}^{\I\phi_a}\equiv r_Rr_L\mathrm{e}^{2\I k_{z,a}L_z}=1,\label{crit_guided}
\end{equation}
which allows to find them in function of frequency, or alternatively, their cosines in function of normalized
frequency $\eta_a\od{\varTheta}$. Because
\begin{equation}
k_{z,a}=\frac{2\pi n_{(0)}\varTheta \eta_a\od{\varTheta}}{\Lambda},
\end{equation}
for a guided mode $\abs{r_L}=\abs{r_R}=1$ and
\begin{equation}
\diff{\phi_a}{\varTheta}=0,
\end{equation}
it is easy to find, that
\begin{equation}
\diff{\eta_a}{\varTheta}=
-\frac{\imag{\frac{1}{r_L r_R}\pdiff{}{\varTheta}\bkt{r_L r_R}}+4\pi\eta_a\frac{\Lambda^{(0)}}{\Lambda}}
{\imag{\frac{1}{r_L r_R}\pdiff{}{\eta}\bkt{r_L r_R}}+4\pi\varTheta\frac{\Lambda^{(0)}}{\Lambda}}.
\end{equation}
This equation can be solved numerically and allows to obtain values of $\eta_a$ and its derivative
for different $\varTheta$ easier, than by solving \eqref{crit_guided} directly (with e.g. bisection method).

Because
\begin{equation}
k_\parallel=\frac{2\pi n_{(0)}\varTheta\sqrt{1-\eta_a^2\od{\varTheta}}}{\Lambda}
\end{equation}
integration over $\vec k_\parallel$ can be replaced by integration over $\varTheta$, which is
carried out immediately, as
\begin{equation}
\Dirac{\Omega_{10}-\omega_{k_a}}=\frac{\Dirac{\varTheta-\varTheta_{10}}}{2\pi f_0},
\end{equation}
and the guided modes' contribution to spontaneous emission rate turns out to be:
\begin{widetext}
\begin{multline}
\Gamma_\GM=\frac{3\pi^2 n_{(0)}^2\locf^2\Gamma^\mathrm{fs}}{2 \varTheta_{10}}
\Bigg[\sum_{a\in\GM_{\varTheta,\TE}}\frac{\dip_x^2}{\dip_{10}^2}\Lambda\abs{P_k}^2\abs{1-\eta_{a,\TE}^2\od{\varTheta}-\varTheta\eta_{a,\TE}\od{\varTheta}\diff{\eta_{a,\TE}\od{\varTheta}}{\varTheta}}
\\+\sum_{a\in\GM_{\varTheta,\TM}}\frac{\dip_x^2\Lambda\abs{Q_{k\parallel}}^2+2\dip_z^2\Lambda\abs{Q_{k\perp}}^2}{\dip_{10}^2}\abs{1-\eta_{a,\TM}^2\od{\varTheta}-\varTheta\eta_{a,\TM}\od{\varTheta}\diff{\eta_{a,\TM}\od{\varTheta}}{\varTheta}}\Bigg]_{\varTheta=\varTheta_{10}},
\end{multline}
\end{widetext}
where $\GM_{\varTheta,\pol}$ denotes the set of guided modes for normalized frequency $\varTheta$
and polarization $\pol$.
For guided modes $P_k$, $Q_{k\parallel}$ and $Q_{k\perp}$ are given by expressions of the same form as in
the case of substrate modes and the relevant definition of $F_{k\pol}$ can be found in \cite{APPA112505}
or \cite{APPA115660} as well. However, it is worth to stress, that in case of guided modes coefficients
$F_{k\pol}$ have the dimension of length, thus ratios $F_{k\pol}/\Lambda$ are dimensionless.
The term $\abs{1-\eta_{a,\pol}^2\od{\varTheta}-\varTheta\eta_{a,\pol}\od{\varTheta}\diff{\eta_{a,\pol}\od{\varTheta}}{\varTheta}}$
is often neglected, e.g. in \cite{Rigneault96}. It is in fact equal to one in a~resonator with metallic mirrors,
but in a multilayer structure it can differ from unity quite significantly.

\section{Exemplary results}\label{sec_results}
\begin{figure}[!tb]
\begin{center}
\includegraphics[width=.45\textwidth]{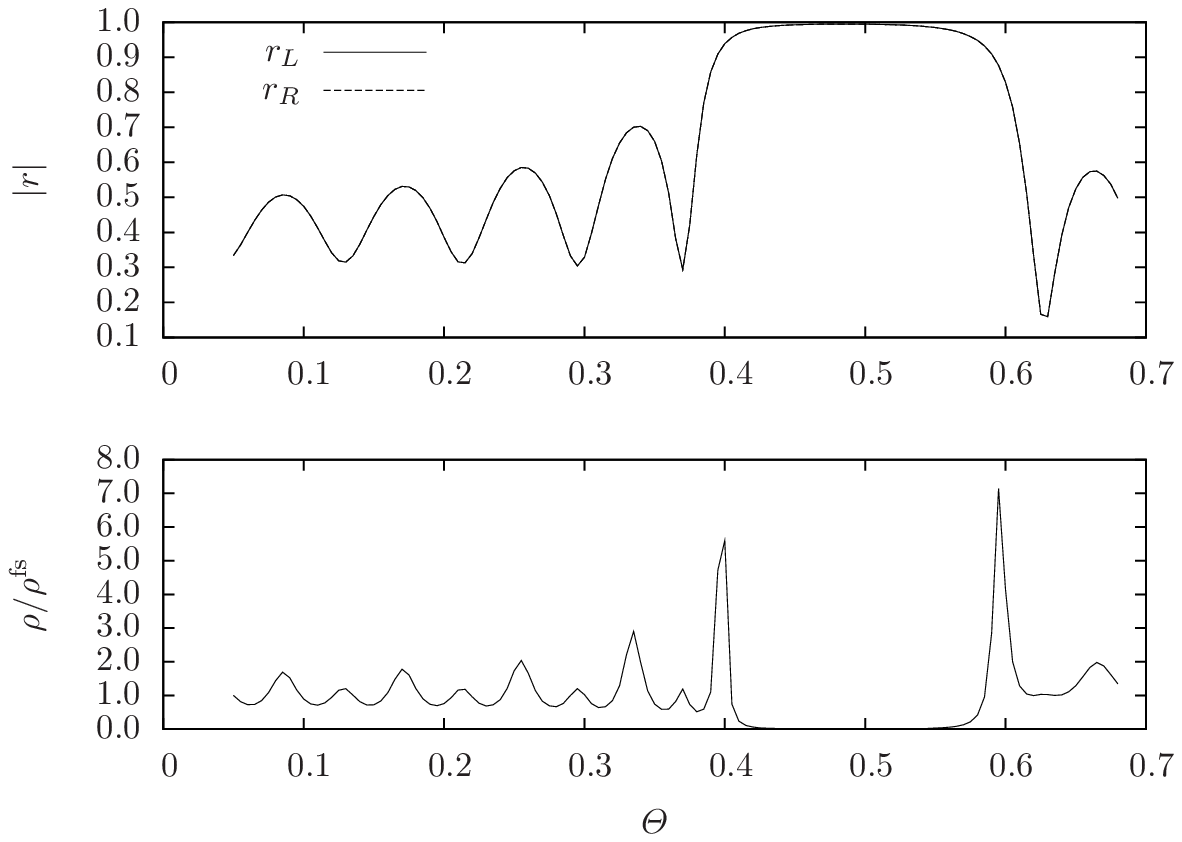}
\includegraphics[width=.45\textwidth]{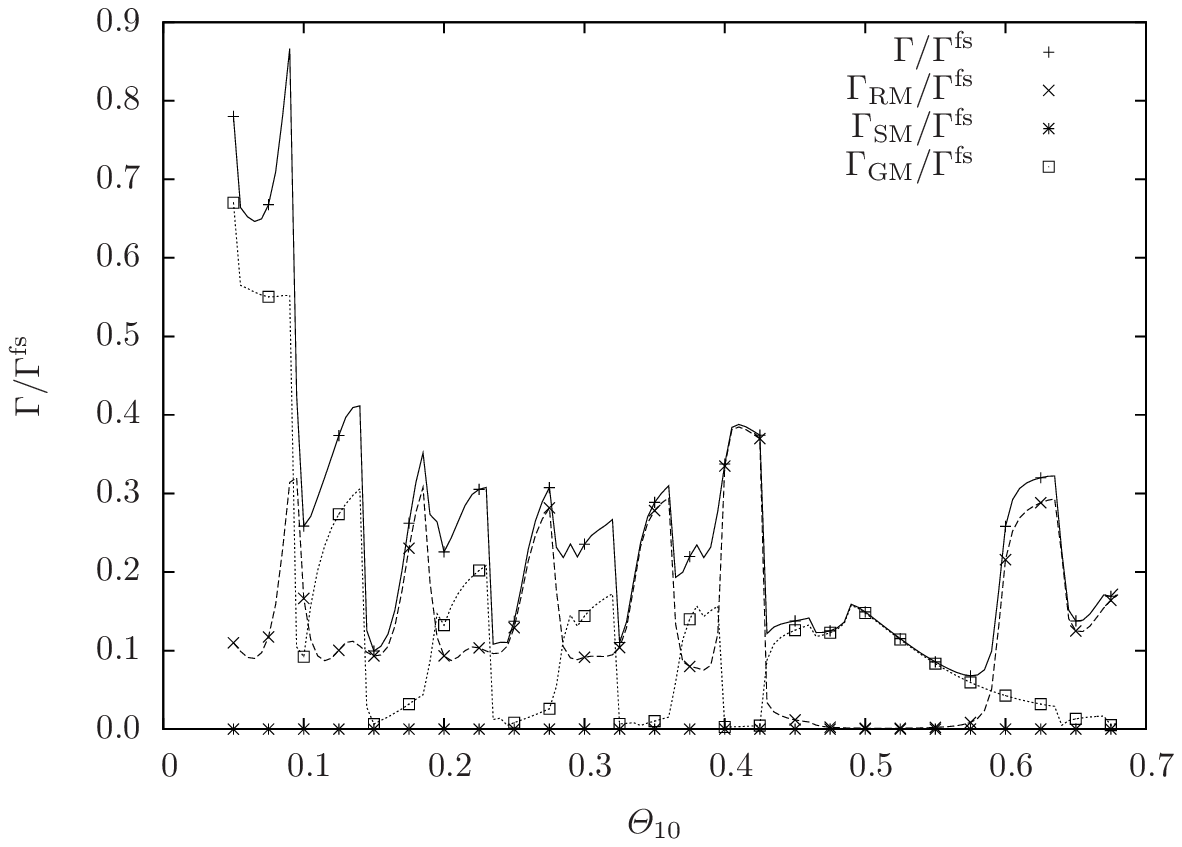}
\caption{Absolute value of reflection coefficients, mode spectrum (both for $\eta=1$) and spontaneous emission rates for structure \str{1}.}\label{fig_S1}
\end{center}
\end{figure}
\begin{figure}[!tb]
\begin{center}
\includegraphics[width=.45\textwidth]{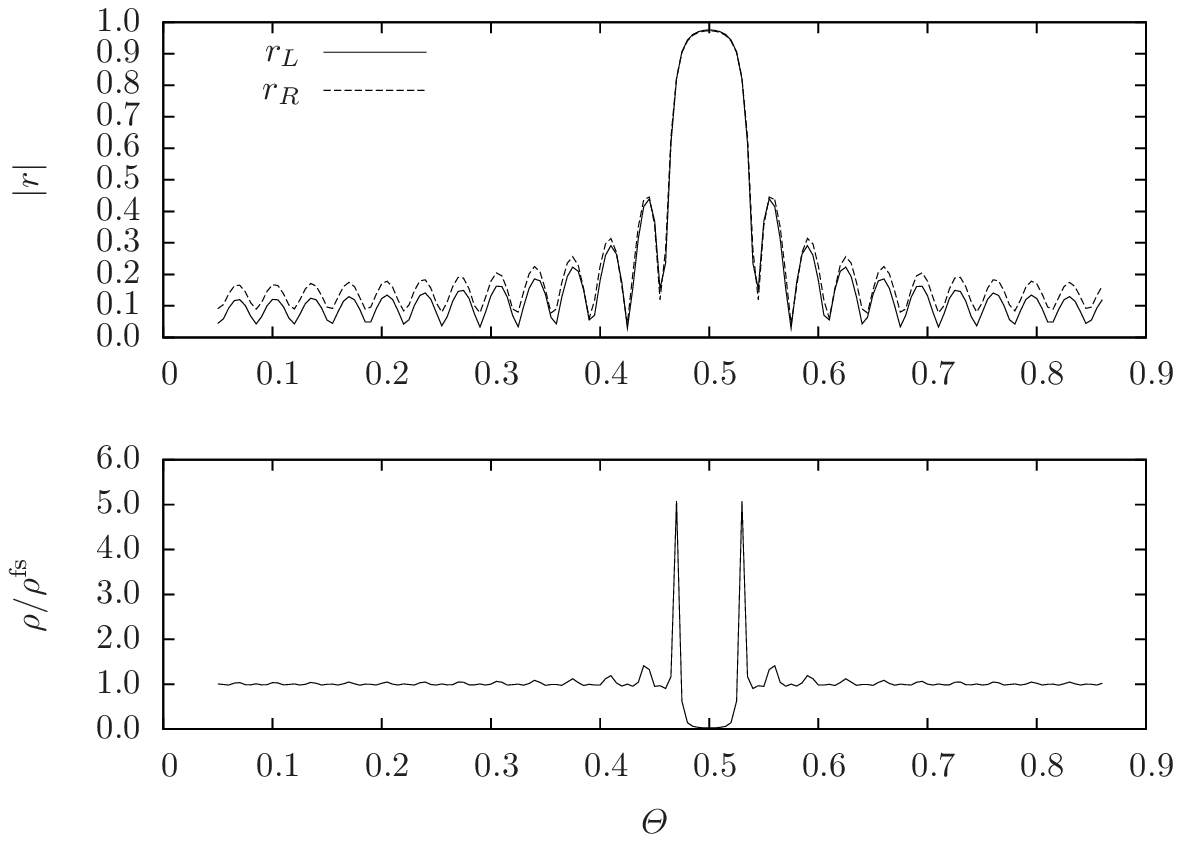}
\includegraphics[width=.45\textwidth]{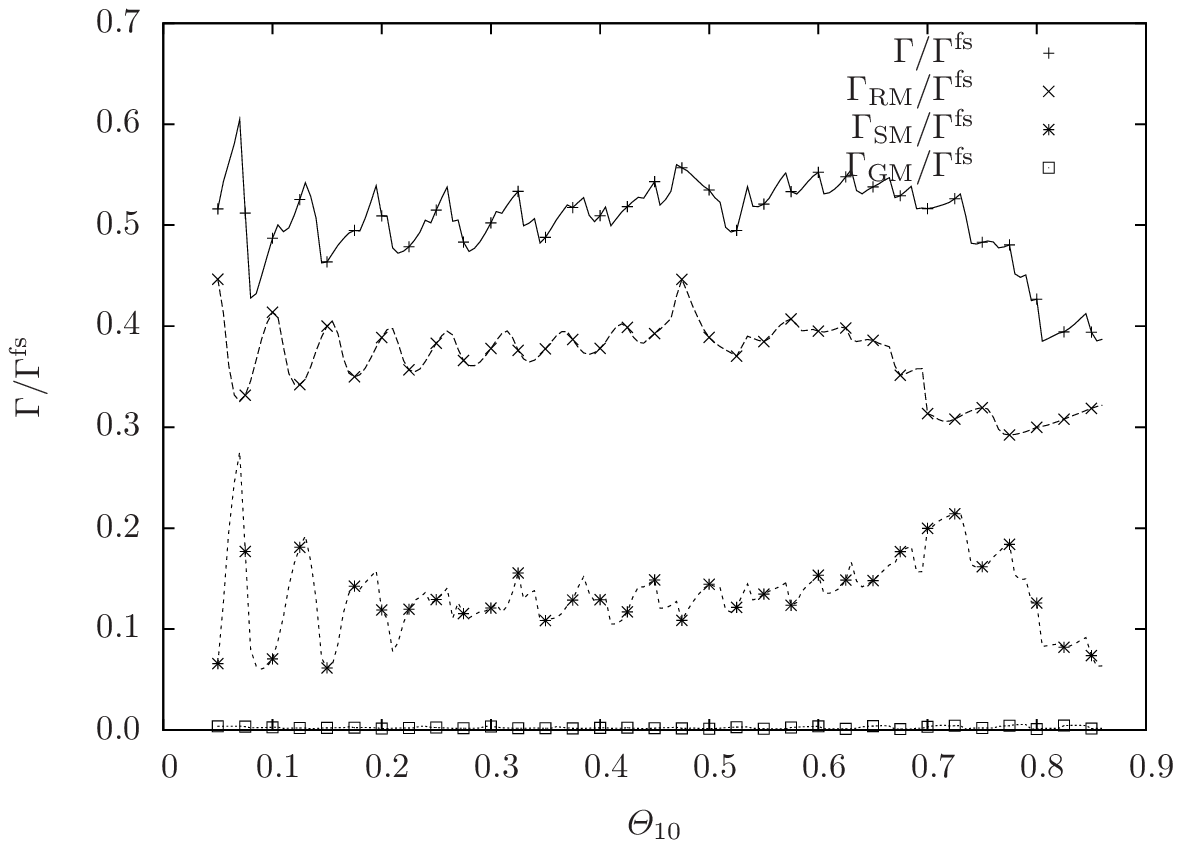}
\caption{Absolute value of reflection coefficients, mode spectrum (both for $\eta=1$) and spontaneous emission rates for structure \str{2}.}\label{fig_S2}
\end{center}
\end{figure}
\begin{figure}[!tb]
\begin{center}
\includegraphics[width=.45\textwidth]{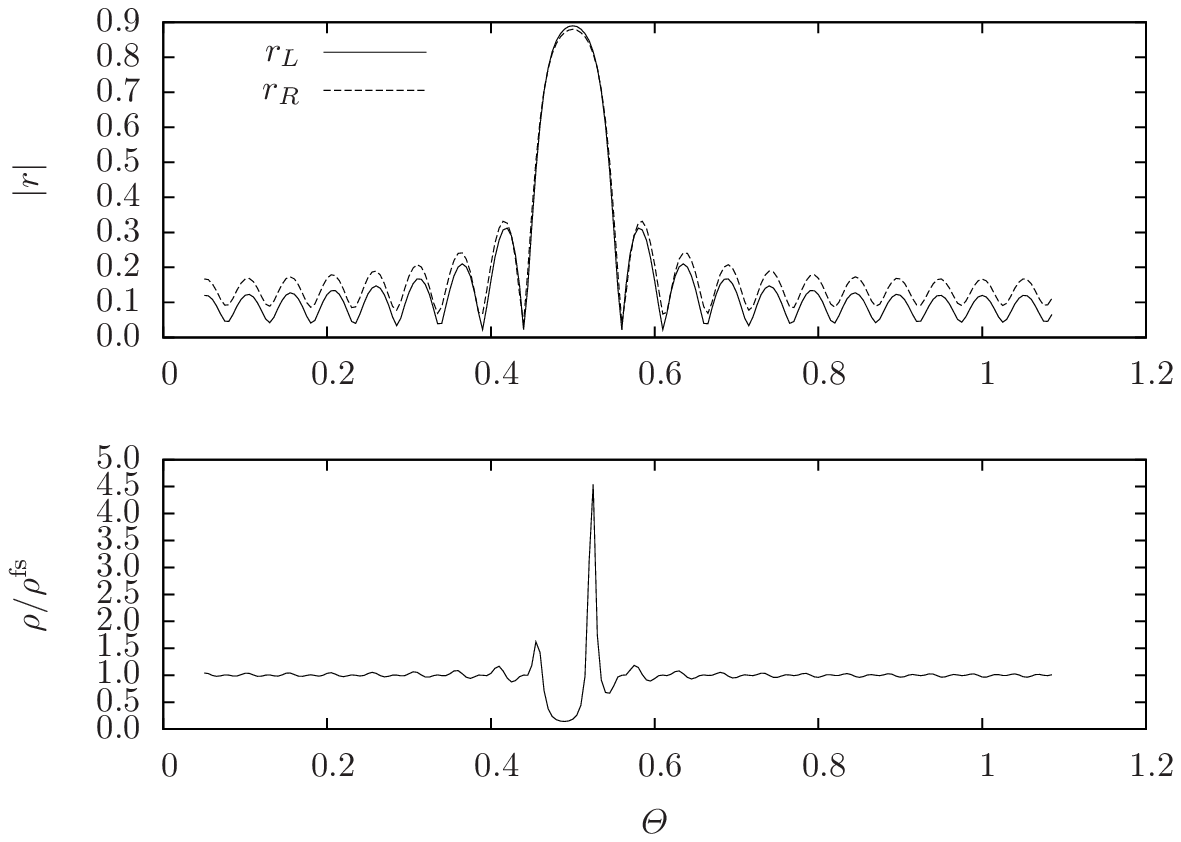}
\includegraphics[width=.45\textwidth]{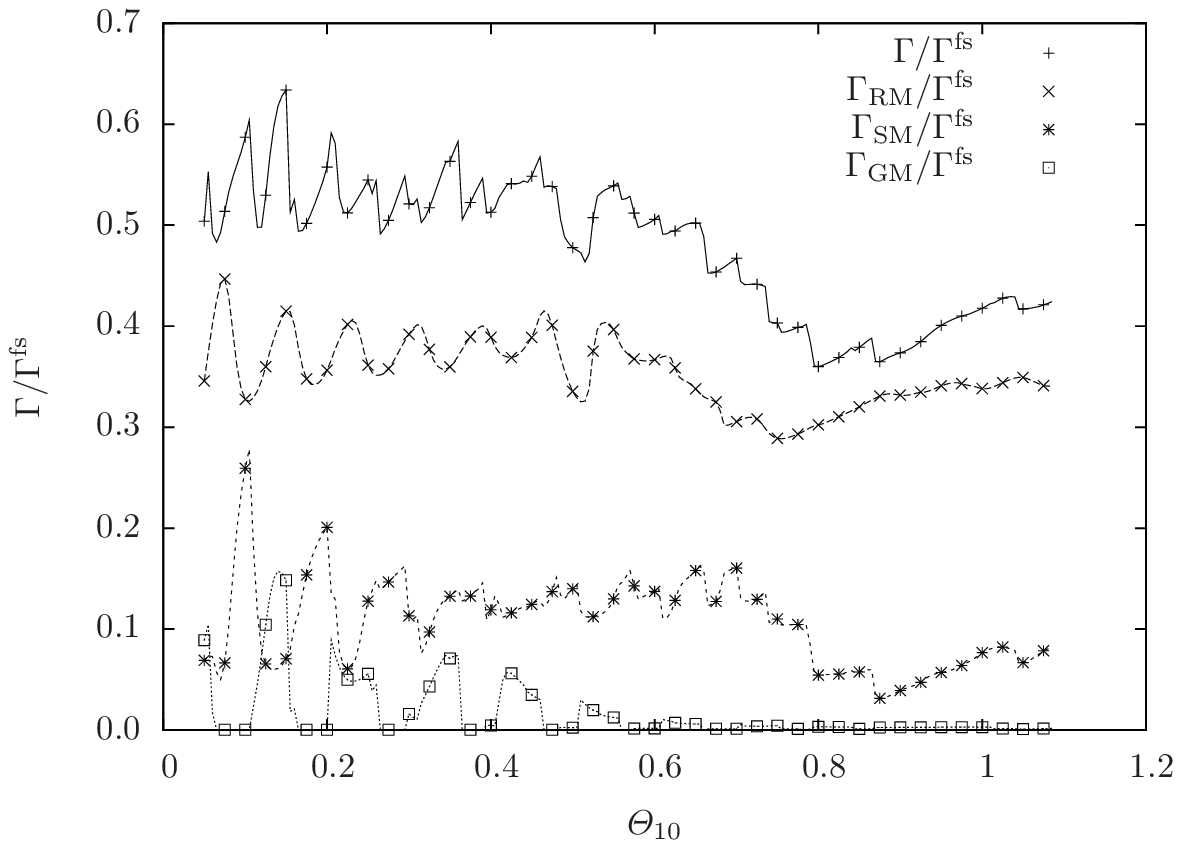}
\caption{Absolute value of reflection coefficients, mode spectrum (both for $\eta=1$) and spontaneous emission rates for structure \str{3}.}\label{fig_S3}
\end{center}
\end{figure}
\begin{figure}[!tb]
\begin{center}
\includegraphics[width=.45\textwidth]{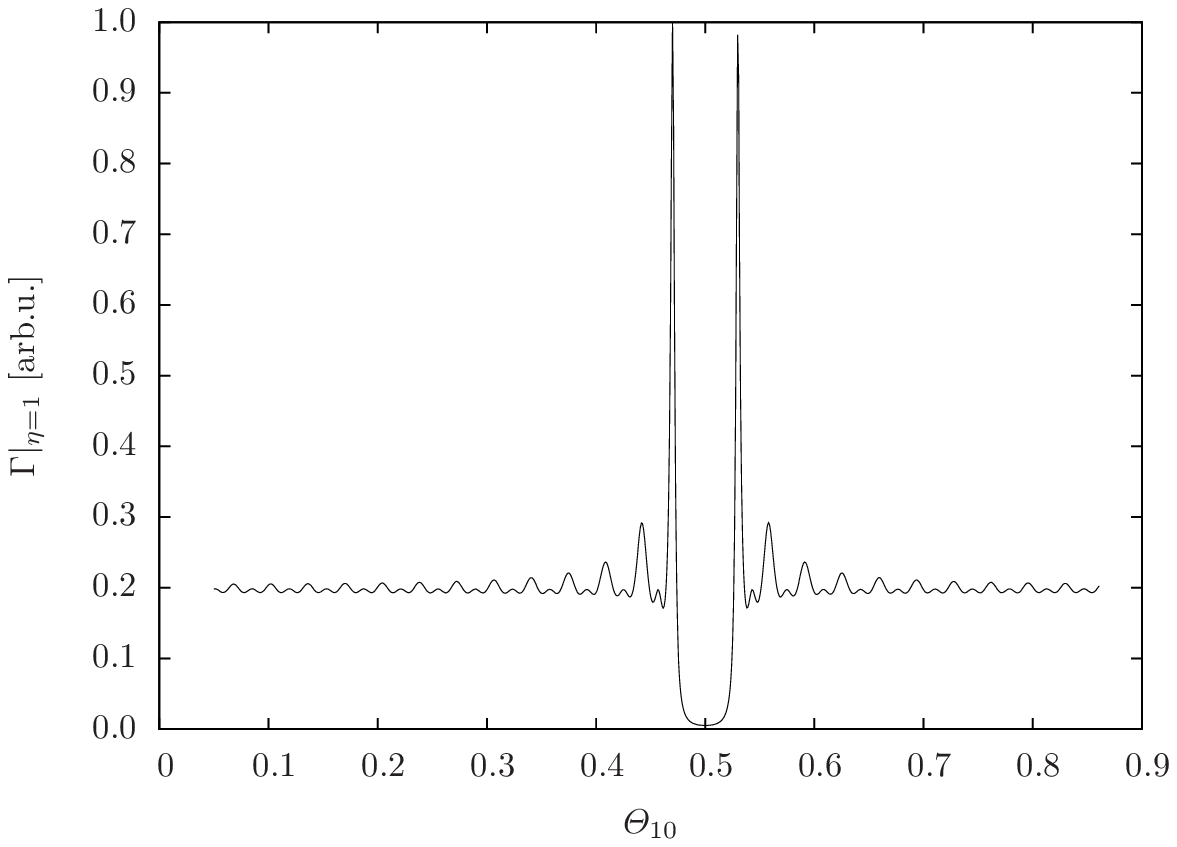}
\caption{Contribution to $\Gamma$ at $\eta=1$ for structure \str{2}.}\label{fig_S2_eta1_Gamma}
\end{center}
\end{figure}
\begin{figure}[!tb]
\begin{center}
\includegraphics[width=.45\textwidth]{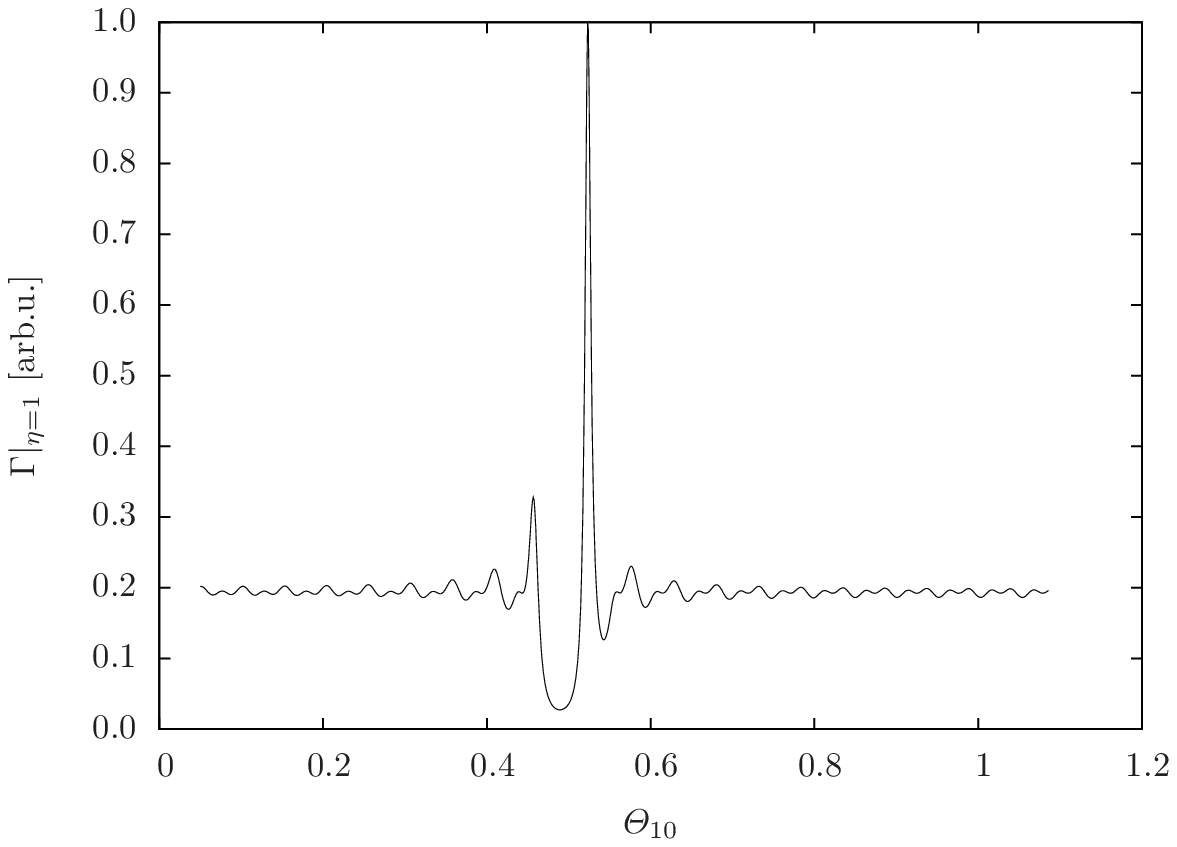}
\caption{Contribution to $\Gamma$ at $\eta=1$ for structure \str{3}.}\label{fig_S3_eta1_Gamma}
\end{center}
\end{figure}
The expression derived in previous sections allow to calculate and present numerical values of
the spontaneous emission rate, provided an explicit expression for $\locf$ is specified.
Following Glauber and Lewenstein \cite{Glauber91}, we put
\begin{equation}
\locf=\frac{3n_{(0)}^2}{2n_{(0)}^2+1},
\end{equation}
what is the proper choice for substitutional impurities \cite{deVries98}. We have conducted calculations
for the following structures:
\begin{itemize}
 \item \str{1}, with $\Lambda_1/\Lambda=0.65$, $n_1=3$, $n_2=1.6$, $n_{(R)}=n_{(L)}=1$ and $N_R=-N_L=12$ (i.e.~$6$ periods in each direction starting the count from the $j=0$ layer);
 \item \str{2}, with $\Lambda_1/\Lambda=0.5$, $n_1=1.4$, $n_2=1.2$, $n_{(R)}=1$, $n_{(L)}=1.1$ and $N_R=-N_L=30$;
 \item \str{3}, with $\Lambda_1/\Lambda=0.5$, width defect in the $j=0$ layer: $\Lambda^{(0)}/\Lambda=0.7$,
 $n_1=1.4$, $n_2=1.2$, $n_{(R)}=1$, $n_{(L)}=1.1$ and $N_R=-N_L=20$ (i.e.~$10$ periods in each direction starting the count from the $j=0$ layer).
\end{itemize}

Results for structure \str{1} are presented in Fig.~\ref{fig_S1}. The top plot contains the absolute
values of reflection coefficients of stacks of layers on both sides of the $j=0$ layer, calculated for
$\eta=1$, i.e.~perpendicular incidence. The middle plot contains mode spectrum (normalized to the
free space value) for the same angle of incidence. These two plots show the band gaps forming in the
photonic crystal. Band gaps appear near the normalized frequency $\varTheta_\mathrm{B}=0.5$, corresponding
to the Bragg wavelength $\lambda_\mathrm{B}=2\Lambda$ and its multiplicities. For a band gap to form,
a high quality factor of the cavity (layer) is required, therefore they appear in the regions with high
reflection, which are characterized by low values of mode spectrum. In perfect photonic crystal, mode
spectrum in a band gap would be equal to zero, but in a real, finite structure it is not possible.
Mode spectrum is high at the edges of a band gap, resembling an effect of the modes being ``pushed out''
of the bang gap. It is also possible to relate values of mode spectrum to profiles of field distribution
-- for high value, the mode has strong field in the layer, while for low values its field in the layer
is weak. Because the emission rate is high for modes exhibiting strong fields at the location of the atom,
modes with high mode spectrum are expected to share the most of emitted power, while those with low mode
spectrum have negligible contributions to the overall rate. These observations allow to interpret
the bottom plot, showing emission rate split into contributions from all types
of modes. Because for structure \str{1} the dielectric outside is the same, there are no substrate modes
and $\Gamma_\SM=0$. The band gap, seen in the middle plot, for a lower $\eta$
would slide towards higher $\varTheta$ (because the $k_z$ component of the wave vector is determined
by $\lambda_\mathrm{B}$), but in this structure band gaps for all $\eta$
partially overlap, thus, it exhibits a full band gap. That is why in the plot of the radiation modes' contribution
$\Gamma_\RM/\Gamma$ there can be seen a region of $\varTheta$, in which the emission is practically forbidden.
However, there is present a contribution from a few guided modes, meaning that for transition frequencies in the band gap
the excitation of an atom would decay quite slowly with emission into these modes.

Similar results for structures \str{2} and \str{3} are plotted in Figs.~\ref{fig_S2}
and~\ref{fig_S3}. For these structures, $n_1$ and $n_2$ are too low for a full band gap
to form, therefore they do not inhibit spontaneous emission as selectively as \str{1}.
Instead of trying to trace characteristic spots in $\Gamma$, which is roughly constant,
it is much more interesting to investigate the contributions to the decay rate from modes
with a given value of $\eta=1$ (the same as for which mode spectrum was plotted),
shown in Figs.~\ref{fig_S2_eta1_Gamma} and~\ref{fig_S3_eta1_Gamma}.
These plots make it evident, that there is a~strict relation between mode spectrum
and contribution to decay rate, and therefore probability of emission into a~particular mode,
given by \eqref{eq_absbkpol2}, as well.
It is also interesting to note, that the defect of structure \str{3}, which results in
a defect mode seen in the mode spectrum inside the band gap, does not increase the total decay
rate, but it rather helps to direct the emission into specific modes.
Because at each frequency defect modes occur only at one angle of incidence,
while maxima at the edges of band gap can appear at two, thus, structure \str{3} makes
a better control of the emission, than \str{2}.

\begin{figure}[!tb]
\begin{center}
\includegraphics[width=.45\textwidth]{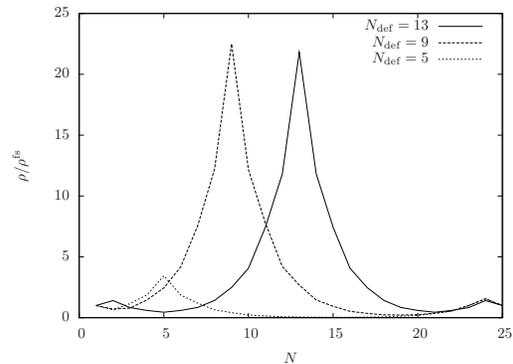}
\caption{Mode spectrum of a defect mode in consecutive $n_1$ layers of photonic crystal \str{1} with a defect introduced in layer $N_\mathrm{def}$ (for $\eta=1$, $N$ stands for the index of elementary cell).}\label{fig_pos}
\end{center}
\end{figure}
Characteristic emission rate for an atom in a specific layer is proportional to mode spectrum in that layer,
therefore analysis of variation of mode spectrum in different layers reveals important information
about the structure. One of interesting results is presented in Fig.~\ref{fig_pos}.
In this figure, there is plotted mode spectrum of a defect mode in $n_1$ layers of different elementary cells
of structure \str{1} with a defect introduced in one of its layers. It is clearly seen, that the defected layer is
characterized by the maximal value of mode spectrum, which becomes quickly much smaller in surrounding layers.
The maximal value depends also on the situation of the defected layer. It follows from obtained expressions,
that the same is the behaviour of spontaneous emission rate into the defect mode. Similarly, in case of a~1D photonic crystal
with multiple defects, analyzed in \cite{OQEL}, the results presented therein can be immediately related with
spontaneous emission rate. This conclusion could not have been drawn from density of states calculated
for the whole structure or with a model assuming external excitation and is a~good example of one of the effective
resonator model's advantages.

\section{Summary}\label{sec_summary}
In this paper we have obtained the expressions describing spontaneous emission rates in a structure
of finite one-dimensional photonic crystal with arbitrary defects. Our derivation has been based on
the effective resonator model, which allows to calculate field distributions and defines a quantity called mode spectrum,
which contains information about physical properties of a~particular layer of the structure.
Thanks to this property mode spectrum is much more useful for analysis of the structure's properties, than e.g. density of states,
characterizing the whole structure, because it allows to easily observe effects associated with different parts
of the structure. We have defined a set of dimensionless parameters
of the structure, which are very convenient to work with, in particular the derived formulas for
the spontaneous emission decay rate are fully expressed by them. We have discussed contributions
from modes of different types and shown, that contributions from radiation and substrate modes
are proportional to mode spectrum. This is an important result, because calculation of mode spectrum
is much easier than of field distribution and provides an easy way of investigation
how a structure affects emission of electromagnetic radiation. Employing mode spectrum
for an analysis is particularly beneficial in case of defected structures, allowing to easily
reveal defect modes or their behaviour throughout the structure, what is not always easy to
accomplish with other models. Exemplary results
included in the text indicate, that the developed description can easily characterize
spontaneous emission from defected layer or any nearby layer, and provides information on
how localization of field changes when the defected layer is moved in the photonic crystal.
Thus, we have shown that the effective resonator model is a tool suitable for
modeling of spontaneous emission from multilayer structures, including defected finite one-dimensional
photonic crystals, which can be easily applied in various practical designs.

\section*{Acknowledgments}
This work has been financially supported by State Committee for Scientific Research, project N N515 052535.

\end{document}